\documentclass[prl,twocolumn,aps,superscriptaddress,letterpaper]{revtex4-1}
\usepackage{graphicx}
\usepackage{amsmath,amssymb}
\usepackage{mathrsfs}
\usepackage[normalem]{ulem}
\usepackage{color}
\usepackage{soul}

\newcommand{\be}{\begin{equation}}
\newcommand{\ee}{\end{equation}}
\newcommand{\bea}{\begin{eqnarray}}
\newcommand{\eea}{\end{eqnarray}}

\newcommand{\vbg}{V_{{\rm BG}}}
\newcommand{\vtg}{V_{{\rm TG}}}
\newcommand{\cbg}{C_{{\rm BG}}}
\newcommand{\ctg}{C_{{\rm TG}}}
\newcommand{\Done}{\Delta_1}
\newcommand{\Dtwo}{\Delta_2}

\newcommand{\MIT}{Department of Physics, Massachusetts Institute of Technology, Cambridge, MA 02139, USA}
\newcommand{\UFMG}{Departamento de Fisica, Universidade Federal de Minas Gerais, Belo Horizonte, MG 31270-901, Brazil}

\newcommand{\NIMS}{Advanced Materials Laboratory, National Institute for Materials Science, 1-1 Namiki, Tsukuba 305-0044, Japan}

\newcommand{\Chula}{Department of Physics, Faculty of Science, Chulalongkorn University, Patumwan, Bangkok 10330, Thailand}
\newcommand{\UCB}{Department of Physics, University of California, Berkeley, California 94720, USA}

\renewcommand{\phi}{\varphi}
\renewcommand{\epsilon}{\varepsilon}

\newcommand{\ignore}[1]{}

\begin{document}

\title{Landau level splittings, phase transitions, and non-uniform charge distribution in trilayer graphene}

\author{Leonardo C. Campos}
\affiliation{\MIT}
\affiliation{\UFMG}

\author{Thiti Taychatanapat}
\affiliation{\Chula}

\author{Maksym Serbyn}
\affiliation{\UCB}

\author{Kawin Surakitbovorn}
\affiliation{\MIT}

\author{Kenji Watanabe}
\affiliation{\NIMS}

\author{Takashi Taniguchi}
\affiliation{\NIMS}

\author{Dmitry A. Abanin}
\affiliation{Department of Theoretical Physics, University of Geneva, 1211 Gen\`{e}ve 4, Switzerland}

\author{Pablo Jarillo-Herrero}
\email[]{pjarillo@mit.edu}
\affiliation{\MIT}
\begin{abstract}

We report on magneto-transport studies of dual-gated, Bernal-stacked trilayer graphene (TLG) encapsulated in boron nitride crystals. We observe a quantum Hall effect staircase which indicates a complete lifting of the twelve-fold degeneracy of the zeroth Landau level. As a function of perpendicular electric field, our data exhibits a sequence of phase transitions between all integer quantum Hall states in the filling factor interval $-8<\nu<0$. We develop a theoretical model and argue that, in contrast to monolayer and bilayer graphene, the observed Landau level splittings and quantum Hall phase transitions can be understood within a single-particle picture, but imply the presence of a charge density imbalance between the inner and outer layers of TLG, even at charge neutrality and zero transverse electric field. Our results indicate the importance of a previously unaccounted band structure parameter which, together with a more accurate estimate of the other tight-binding parameters, results in a significantly improved determination of the electronic and Landau level structure of TLG.

\end{abstract}

\maketitle

The electronic properties of ABA-stacked TLG are being intensively investigated~\cite{Craciun2009a,Taychatanapat2011,Bao2011,Henriksen2012,Zou2013,Lee2013} due to its distinct band structure which consists of two overlapping monolayer graphene-like (MLG) and bilayer graphene-like (BLG) bands~\cite{Latil2006,Partoens2006,Guinea2006,McCann2006prl,Aoki2007,Koshino2009,Lui2009,Craciun2009a,Koshino2010}. However, in contrast to MLG and BLG, which are gapless, both subbands in ABA-stacked TLG  are gapped, with small masses of the order of a few meV.

\begin{figure}[ht]
  \includegraphics[width=3.4in]{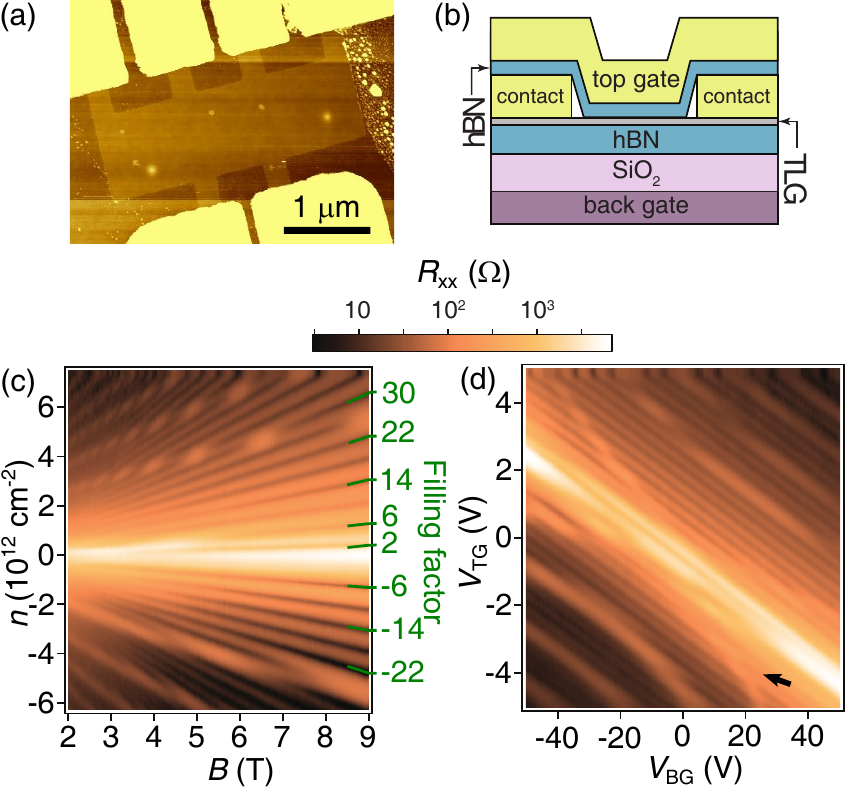}
  \caption{\ignore{\textbf{Device fabrication and quantum Hall effect}}
(a) AFM image of device 1 four probe Hall bar geometry before top gate preparation. The length and width of device 1 are 3~$\mu$m and 1~$\mu$m, respectively.
(b) Schematic cartoon of the sample showing the configuration of top gate and back gate electrodes.
(c) $R_{xx}$ as a function of $n$ and $B$ at $D = 0$.
(d) $R_{xx}$ as a function of top gate and back gate voltages at $B = 4$~T. The black arrow points to the quickly dispersing LL crossings at $D \neq 0$ on the electron side. These LL crossings also occur on the hole side. Data in (c, d) are measured at $4$~K.
  }
  \label{f1}
\end{figure}

One of the most interesting characteristics of ABA-stacked TLG compared to other graphene systems is the way in which its band structure is modified by a perpendicular electric field~\cite{Koshino2009,Koshino2009b,Koshino2010,Jhang2011,Taychatanapat2011,Campos2012,Serbyn2013}. Theory predicts that a weak electric field hybridizes the MLG-like and BLG-like bands, rather than inducing a band gap, as in BLG or ABC-stacked TLG~\cite{Bao2011,Jhang2011,Lui2011,Khodkov2012,Khodkov2015}. The hybridized bands are characterized by a strong trigonal warping. For very strong electric fields, a new set of Dirac points was theoretically predicted, with masses and velocities that are controlled by the electric field~\cite{Serbyn2013,Morimoto2013}. Thus {\it biased} TLG can potentially provide an opportunity to study chiral carriers with tunable anisotropic dispersion, different symmetry and higher valley degeneracy (6 as opposed to two in MLG and BLG), not accessible in MLG and BLG.

Here we report on transport studies of high-mobility TLG samples in the quantum Hall effect (QHE) regime. We fabricated dual-gated TLG samples encapsulated in hexagonal boron nitride~\cite{Dean2010,Mayorov2011} crystals (hBN) (Figs.~1(a)-(b)), which allowed us to independently control the carrier density $n$ and perpendicular electric displacement field $D$. We use magneto-transport measurements to study how Landau levels (LLs) evolve under $D$. By inspecting the pattern of LL crossings resulting from the hybridization of the BLG-like and MLG-like bands as a function of $n$, $D$, and magnetic field $B$, we are able to refine the values of the TLG band structure parameters.

\begin{figure}
  \includegraphics[width=3.4in]{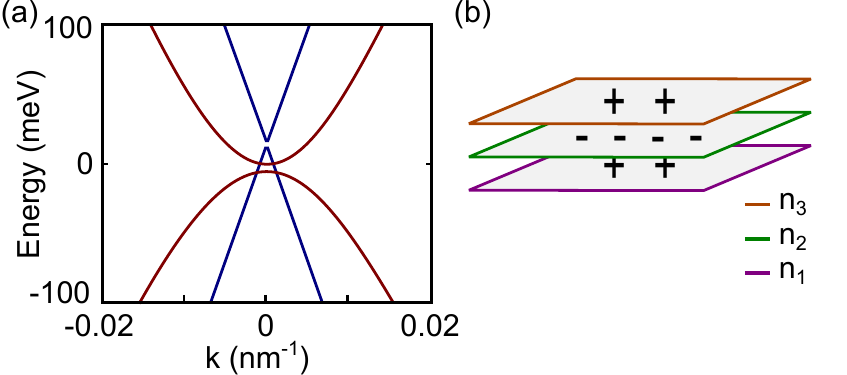}
  \caption{\ignore{\textbf{Band structure at low energy and charge distribution at $n$ = 0}}
(a) Band structure of TLG for $\gamma_0 = 3.1$~eV, $\gamma_1 = 0.39$~meV, $\gamma_2=-18$~meV, $\gamma_3=315$~meV, $\gamma_4=100$~meV, $\gamma_5=10$~meV, $\delta = 15$~meV, $\Done = 0$ and $\Dtwo = 1.8$ meV.
(b) Schematic cartoon showing charge distribution at $\Done = 0$ and $n = 0$.
  }\label{f2}
\end{figure}

Additionally, we observe a quantum Hall staircase with plateaus spaced by $e^2/h$, indicating a complete lifting of the low-lying LL degeneracies~\cite{Nomura2006,Alicea2006,Herbut2007,Bao2010}. We find a number of $D$-driven phase transitions between different integer QH states within the zeroth LL. The phase transitions occur at every integer filing factor in the interval $-8 \leq \nu \leq 0$. While in MLG and BLG the zeroth LL splittings usually arise from Coulomb interaction effects, we argue that the splittings and phase transitions that we observe are consistent with a single-particle picture.  We develop a theoretical model and show that the observed phase transitions can be attributed to multiple LL crossings induced by the transformation of the TLG band structure under $D$. However, the sequence of phase transitions implies a finite value for a previously neglected parameter in the band structure, which corresponds to the difference of electrostatic potential of the middle layer from the average potential on outer layers of TLG. Such a term, allowed by symmetry, points to a non-uniform internal electric field in TLG, which is caused by {\it non-zero charge density} on the middle layer. In particular, even at the charge neutrality and zero $D$ there are non-zero charges on outer and inner layers~\cite{Koshino2009}, see Fig.~\ref{f2}.

Our devices are fabricated using hBN crystals as a local substrate to improve device quality~\cite{Dean2010}. Then another hBN flake is transferred on top of TLG to cover the sample  [Fig.~\ref{f1}(a)-(b)]~\cite{Campos2012}. This allows us to fabricate a top gate, enabling the independent control of both $n$ and $D$. These two quantities can be related to the voltages on the top and bottom gates, $\vtg$, $\vbg$, via $n = (\ctg \vtg + \cbg \vbg)/e$ and $D = (\ctg \vtg - \cbg \vbg)/2$, where $\ctg,\cbg$ are the top gate and back gate capacitances per unit of area. We have measured two TLG devices, one in a Hall bar geometry (device 1) and the other in a two-terminal geometry (device 2).

To establish that our graphene devices indeed consist of ABA-stacked TLG, we first studied magneto-transport at $D = 0$, where the MLG-like and BLG-like bands are not hybridized. The longitudinal resistance $R_{xx}$, shown in Fig.~\ref{f1}(c), exhibits minima at many integer filling factors, indicating the high quality of the device. More importantly, characteristic crossings between MLG-like and BLG-like LLs are evident, which originate from the different scaling of LL energies with $B$: ignoring small mass terms, LL energies scale as $\sqrt{B}$ and $B$ for MLG-like and BLG-like bands, respectively~\cite{Taychatanapat2011,Novoselov2005,Zhang2005n,McCann2006prl,Guinea2006,Barlas2012,Zhang2012}.

The pattern of LL crossings in Fig.~\ref{f1}(c) is qualitatively similar to that observed in Ref.~\cite{Taychatanapat2011}, but there are two new important features. First, at high electron density, the locations of LL crossings at fixed $B$ are different, with the crossing points shifted up by one (four-fold degenerate) LL compared to Ref.~\cite{Taychatanapat2011}. For example, at $B = 9$ T, we observe the LL crossing point at filling factor $\nu = 26$ instead of $\nu = 22$ as it was reported in Ref.~\cite{Taychatanapat2011}. We note that, at high hole density, $n <- 2 \times 10^{12}$ cm$^{-2}$, values of $n$ and $B$ at which LL crossings occur are very similar to those in Ref.~\cite{Taychatanapat2011}. The difference in the LL crossing locations on the electron side is due to the different electrostatic conditions of experiment~\cite{Taychatanapat2011}, which was performed at small $D(n)\neq 0$. The second and more important difference occurs at low density. In particular, it is evident that the zeroth LL (which, in the simplified tight-binding model, is 12-fold degenerate) is partially split, and this splitting breaks electron-hole symmetry. This can be seen in Fig.~\ref{f1}(c) for $4\, {\rm T}< B<9 \,{\rm T}$: there is a clear minimum of $R_{xx}$ for filling factor $\nu = +2$ but not for $\nu = -2$. These new features at low and high charge density indicate that the TLG tight binding parameters have to be reevaluated. In particular, we will see below that a non-zero parameter $\Delta_2$, describing the charge redistribution between the central and outer layers, is necessary to explain the electron-hole asymmetry observed at filling factor $\nu=\pm 2$.

Next, we study magneto-transport at $D\neq 0$. The behavior of $R_{\rm xx}$ in device 1 as a function of $V_{\rm TG}$ and $V_{\rm BG}$, illustrated in Fig.~\ref{f1}(d), reveals a series of LL crossings which occur as a function of bias at fixed $B$. Theoretically, such crossings are expected as a result of the LL structure modification induced by the hybridization of MLG-like and BLG-like bands. Notice that LL crossings are visible on both electron and hole sides.

\begin{figure}
  \includegraphics[width=3.4in]{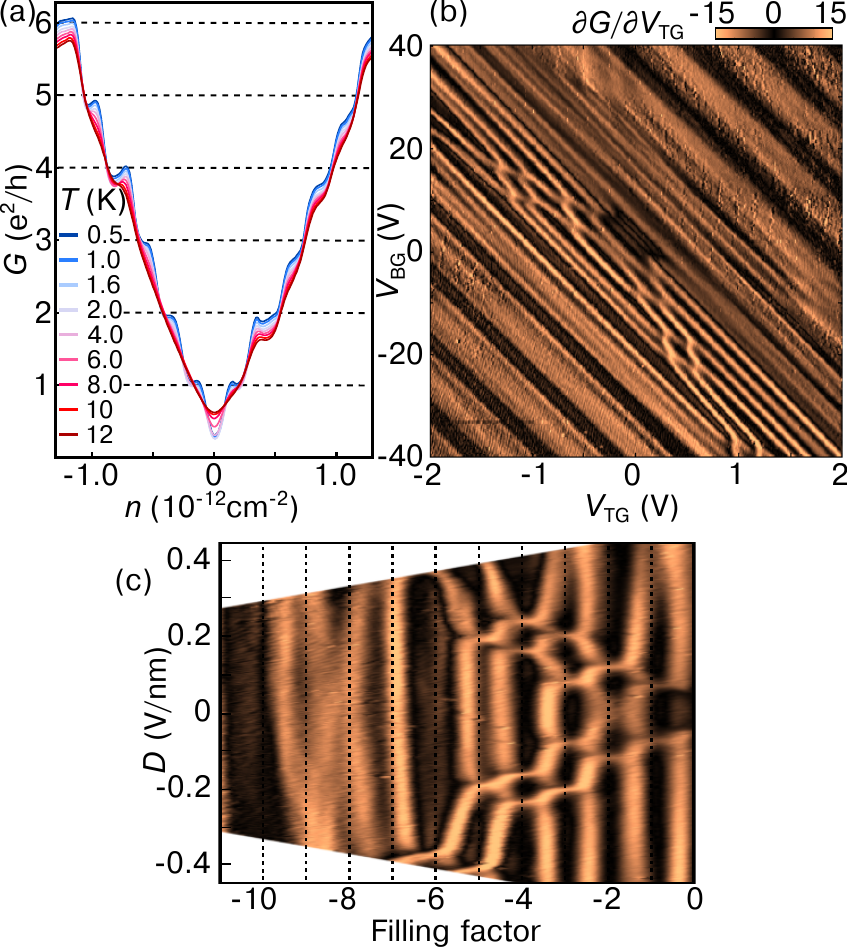}
  \caption{\ignore{\textbf{Broken symmetry states and Landau level (LL) crossings as function of $D$}}
(a) Conductance plateaus at $B$ = 9 T as function of charge density at various temperature. The length and width of device 2 are 3~$\mu$m and 4~$\mu$m, respectively.
(b) Numerical derivative of the conductance at $B$ = 9 T.
(c) Zoom in data shown in (b), showing the crossing of LLs, plotted vs filling factor, $\nu$ and $D \cdot \epsilon_{0}$. Where $\epsilon_0$ is the vacuum permittivity. Data in (b, c) are measured at $300$~mK.
  }\label{f3}
\end{figure}

Since disorder-induced LL broadening (most likely due to charge inhomogeneity) prevents us from resolving the fine structure of LL crossings as a function of $D$, we turn to device 2, which had a much smaller charge inhomogeneity.  The two-terminal conductance of device 2 at $B=9\, {\rm T}$ and $D=0$ shows a QHE staircase with plateau spacing $e^2/h$, indicating a complete splitting of the zeroth and first LL (Fig.~\ref{f3}(a)). It is important to note that the plateaus are better developed on the hole side, where a complete splitting for $-10 \leq \nu < 0$ is visible. On the electron side, only a partial splitting is found, and the plateaus are not as pronounced, which we believe to be the result of electron-hole asymmetric disorder effects.

Next, we explored the behavior of split LLs at finite $D$. Figure \ref{f3}(b) shows differential conductance $dG/dV_{\rm TG}$ as a function of top gate and back gate voltages. It is convenient to re-plot differential conductance as a function of $D$ and filling factor $\nu$ [Fig.~\ref{f3}(c)]. This plot reveals symmetrical crossings at all filling factors in the interval $-8<\nu \leq -1$ for positive and negative $D$. Clearly, as $|D|$ is increased, a pair of LLs quickly shifts away from zero energy, crossing other LLs. This results in a series of QH phase transitions. Note that similar experiment has been performed by Lee \emph{et al.} ~\cite{Lee2013} but the quantum Hall staircase details were obscured by charge inhomogeneity. This behavior is also observed in device 1 -- see Fig.~\ref{f1}(d), where the quickly shifting pair of LLs (which are merged together due to disorder broadening) is pointed by a black arrow. Besides, in device 2, there might be a crossing at $\nu=0$ at very small $D$, but we were not able to fully resolve it. We leave the nature of $\nu=0$ and its transformations with $D$ for future studies.

To understand the LL crossings above, we calculate the {\it single-particle} LL spectrum as a function of $D$ and $B$ numerically, and show that theory captures all the key observed features -- locations of LL crossings, splitting of the zeroth LL, as well as phase transitions between different QH states. We argue that in order to faithfully describe the data, it is necessary to include a term $\Delta_2$ in the tight-binding model of TLG, and moreover, we refine the values of the other tight-binding parameters.

To describe the band structure of trilayer graphene, we use the Slonczewski-Weiss-McClure parametrization of the tight-binding model~\cite{Koshino2009,Serbyn2013}, which includes seven parameters ($\gamma_0$, $\ldots$, $\gamma_5$, $\delta$). Also, to describe the effect of $D$, we introduce the on-site potentials on the three TLG layers, $U_1, U_2, U_3$. Since only potential differences are meaningful, it is convenient to define parameters
$$
\Delta_1=(-e)(U_1-U_3)/2, \,\, \Delta_2=(-e)(U_1-2U_2+U_3)/6.
$$
Parameter $\Delta_1$ describes the potential difference between the two outer layers, and it is zero at $D=0$. However, parameter $\Delta_2$, describing the potential difference between the central layer and the average of the outer layers, is allowed by symmetry even at $D=0$. The presence of $\Delta_2\neq 0$ without displacement field indicates non-zero charge on the middle layer, which is obviously disfavored by electrostatic energy, and arises from the interplay between band structure and electrostatics.

We can confirm the existence of $\Dtwo$ at low density by considering the Landau gaps at $\nu = +2$ and $\nu = -2$.  These two gaps are given by $\Delta \varepsilon_{2} = +3\Dtwo-\gamma_2/2$ and $\Delta \varepsilon_{-2} = -3\Dtwo-\gamma_2/2$, where $\gamma_2$ is the tight binding parameter associated with the hopping integral between A sites in the outer layers ($A_1\leftrightarrow A_3:\gamma_2$)~\cite{Koshino2009,Serbyn2013}. In device 1, we observe a splitting at $\nu = 2$ but not at $\nu = -2$ [Fig.~\ref{f1}(c)]. This implies that $\Dtwo$ is greater than zero or, equivalently, the gap at $\nu = 2$ is larger than that at $\nu = -2$. We note that the gap at $\nu = -2$ is non-zero. However, we do not observe the splitting at $\nu=-2$ in device 1 due to disorder broadening. In device 2 which has a better quality than device 1, we are able to resolve the gap at $\nu = -2$.

We also find the values of the tight-binding parameters which best reproduce the locations of LL crossings at $D = 0$ [Fig.~\ref{f1}(c)]. Parameter $\Delta_2$ is expected to depend on the charge density, since additional charge can also distribute asymmetrically among central and outer layers. Therefore, in our analysis  $\Delta_2$ needs to be reevaluated for each density of charge, to describe electron/hole asymmetries at low charge density (see more details in Supplemental Material~\cite{RefSI}). The other parameters are taken from fitting of the crossing patterns at high charge density, see Table~\ref{table1}.

Next, we explain the LL splittings and fast diving of two LLs away from zero energy with increasing $D$ observed in Fig.~\ref{f3}~(c). The Landau levels are shown in Fig. 4(a), where MLG-like bands are in red lines and BLG-like bands in blue lines. First let us discuss the splitting of the zeroth LL at $D=0$. Ignoring the spin degree of freedom, the zeroth LL consists of six sub-levels. Two sub-levels originate from the MLG-like band, labeled $0_{\pm}^{\rm m}$, and have the highest (and positive) energies (see Fig. 4(b)):
$$
E_{0_{+}^{\rm m}}=\delta-\frac{1}{2} \gamma_5+\Delta_2, \,\, E_{0_{-}^{\rm m}}=-\frac 12 \gamma_2+\Delta_2,
$$
where $\pm $ labels the two valleys.  The four remaining sub-levels originate from the BLG-like band [four blue lines closest to zero energy in Fig. 4(b)], with energies
\be\label{eq:energy1}
E_{0_{+}^{\rm b}}\approx E_{1_{+}^{\rm b}} \approx  -2\Delta_2+\delta E_{01}^+,
\ee
\be\label{eq:energy2}
E_{0_{-}^{\rm b}}\approx E_{1_{-}^{\rm b}} \approx \frac{1}{2}\gamma_2 +\Delta_2+\delta E_{01}^-,
\ee
where $\delta E_{01}^{\pm}$ denotes small single-particle splittings between the $0,1$ sub-levels within the corresponding valley~\cite{Serbyn2013}.
The spin degeneracy is lifted by the Zeeman interaction, $E_Z=g\mu_B B\approx 10\, {\rm K}$ at $B=10\, {\rm T}$. Thus, the zeroth LL is fully split. The gaps estimated by numerical solutions are of the order of a few meV, except for the gap at $\nu=2$ (all BLG-sublevels filled) $\Delta E \sim {15}$ meV. The size of the single-particle gaps is consistent with the observed temperature scale at which the QH states in the interval $0<|\nu|<6$ disappear (T $\sim$ 10 K).

\begin{figure}
  \includegraphics[width=3.4in]{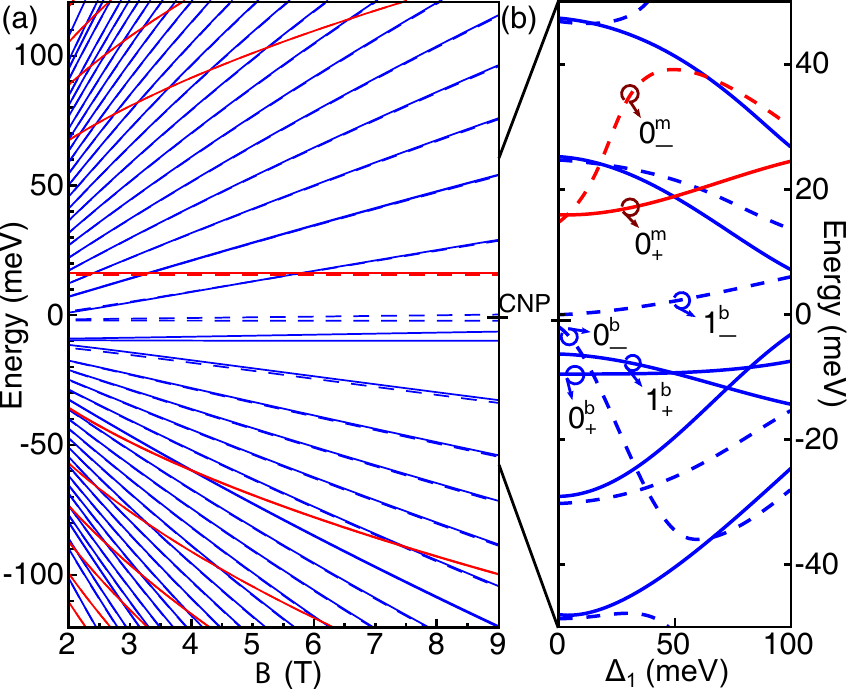}
  \caption{\ignore{\textbf{Simulations of broken symmetry states and Landau level (LL) crossings as function of $\Done$}}
(a) Numerical simulation of the LL spectrum. Red lines are solutions for MLG-like LLs, solid lines address valley $\rm{K}$ and dashed lines address valley $\rm{K_-}$. Blue lines are solutions for BLG-like LLs, solid lines address valley $\rm{K}$ and dashed lines address valley $\rm{K_-}$.
(b) LL crossing as function average inter-layer potential difference $\Done$.
  }\label{f4}
\end{figure}

Below we focus on the region $-6<\nu<0$, where $D$-induced phase transitions occur. Filling factor $\nu=0$ corresponds to filling 6 sub-levels in the BLG-like sector. When the displacement field is applied, the BLG-like and MLG-like bands can no longer be treated as independent, they hybridize, and the effect of trigonal warping increases. These band structure changes lead to a re-arrangement of the Landau sub-levels. In the interval of $D$ values corresponding to $0<|\Delta_1|<30\, {\rm meV}$, achieved in our experiment, the main effect is that the $0_{-}^b$ LL [dashed blue line just below CNP in Fig~\ref{f4}(b)] rapidly sinks down in energy, while LL $1_-^b$ slowly shifts up in energy. Landau levels $0_{+}^b, 1_{+}^b$ disperse rather weakly.

Thus, the phase transitions observed for $-6 < \nu < 0$ can be understood as the result of the spin-split $0_{-}^b$ LL crossing the spin-split $0_+^b,1_+^b$ LLs. This implies that at $D=0$, $E_{0_-^b}>\mathrm{max\,}(E_{0_+^b},E_{1_+^b})$, which puts a limitation on the value of $\Delta_2$. From Eqs.(\ref{eq:energy1},\ref{eq:energy2}), we obtain an estimate
$$
\Delta_2> -\gamma_2/6 \approx 2.7 {\rm \,meV}
$$
\ignore{$$
\Delta_2< -\gamma_2/6\approx ...
$$}
The phase transition observed at $\nu=-1$ is attributed to the crossing between $0_{-}^b \downarrow$ and $1_{-}^b\uparrow$ levels (the Zeeman interaction is very likely strong enough to bring the $0_{-}^b \downarrow$ level above $1_{-}^b\uparrow$ one at $D=0$). Upon increasing $D$, these levels cross. The evolution of LLs and their crossings are shown in Fig.~\ref{f4}(b) for $\Dtwo = 5.7$ meV.

 Our results indicate that parameter $\Delta_2$ is non-zero, implying a non-uniform electric field distribution and presence of charge on the middle layer of TLG. It is interesting to note that, despite the large impact of $\Delta_2$ on the LL spectrum, its effect on the band structure itself at $B=0$ is very small. Indeed, the highly degenerate LL with wave functions often localized on particular layers~\cite{Serbyn2013} are more sensitive to $\Delta_2$ and can lead to its enhancement within a self-consistent picture. On the other hand, in the continuous band structure, the effect of $\Delta_2$ is barely noticeable. To illustrate this, in Fig.~\ref{f2}(a) we plot the energy dispersion at $B = 0$ and $D = 0$ for $\Delta_2 = 1.8$ meV  (value of $\Delta_2$ at $n$ = 0 and $B$ = 9 T). Fig.~\ref{f2}(b) schematically illustrates the non-uniform charge distribution. Essentially, the band structure looks qualitatively the same as that previously reported~\cite{Koshino2009,Taychatanapat2011,Campos2012,Zou2013} at $\Delta_2=0$ -- there is an energy band overlap between MLG-like and BLG-like, and the band gap between bilayer subbands is of the order of $\sim6$ meV at $n=0$. Thus, we need magneto-transport measurements at high field to obtain the value of $\Delta_2$.

\begin{table}
\caption{Tight-binding parameters obtained by fitting the pattern of Landau level crossings shown in Fig.~\ref{f1}(c)}\label{table1}
\begin{center}
    \begin{tabular}{| c | c | c | c | c |c | c |}
    \multicolumn{7}{c}{} \\
    \hline
      $\gamma_0$ & $\gamma_1$ & $\gamma_2$ & $\gamma_3$ & $\gamma_4$  & $\gamma_5$ & $\delta$  \\
      (eV) & (eV) & (meV) & (meV) & (meV) & (meV) & (meV) \\
      \hline
     3.1 & 0.39 & -20 to -16 & 315 & 40 to 140 & 5 to 15 & 12 to 18 \\ \hline
    \end{tabular}
\end{center}
\end{table}

In summary, we have measured and calculated the fine structure of Landau levels in ABA-stacked, dual-gated trilayer graphene devices.
We found full splitting of the zeroth LL, as well as a series of quantum Hall phase transitions induced by electric displacement field. We showed that this behaviour is well-described by a single-particle model, once a previously neglected parameter is included in the tight-binding model of TLG. The existence of this parameter implies a non-trivial electric field distribution between the middle and outer layers in TLG caused by the interplay between electrostatic energy and band structure.

Finally, although our findings are consistent with a single-particle model, there is little doubt that TLG should host interesting fractional quantum Hall states, which should become observable in devices with even lower disorder than those presented here. Our studies show that TLG is a material in which LL energies and wave functions are widely tunable by $D$. In the future it would  be interesting to study whether this tunability could be used to control effective interactions and tuning fractional quantum Hall states~\cite{Papic2011}.

\textit{Acknowledgements}. L.C.C and T. Taychatanapat contributed equally to this work. This work has been primarily supported by the National Science Foundation (DMR-1405221) for device fabrication and transport, and partly by ONR Young Investigator Award N00014-13-1-0610 for data analysis. L.C.C. acknowledges partial support by the Brazilian agency CNPq. M.S. acknowledges support by the Gordon and Betty Moore Foundation's EPiQS Initiative through Grant GBMF4307. D.A.A. acknowledges support by NSERC and the Sloan Foundation. This work made use of the Materials Research Science and Engineering Center Shared Experimental Facilities supported by the National Science Foundation (DMR-0819762) and of Harvard’s Center for Nanoscale Systems, supported by the NSF (ECS-0335765).




\bibliographystyle{apsrev4-1}

\end{document}


\begin{center}
\textbf{Supplemental Material}\\

\large{\textbf{Landau level splittings, phase transitions, and non-uniform charge distribution in trilayer graphene}}
\end{center}

\author{Leonardo C. Campos}
\affiliation{\MIT}
\affiliation{\UFMG}

\author{Thiti Taychatanapat}
\affiliation{\Chula}

\author{Maksym Serbyn}
\affiliation{\UCB}

\author{Kawin Surakitbovorn}
\affiliation{\MIT}

\author{Kenji Watanabe}
\affiliation{\NIMS}

\author{Takashi Taniguchi}
\affiliation{\NIMS}

\author{Dmitry A. Abanin}
\affiliation{Department of Theoretical Physics, University of Geneva, 1211 Gen\`{e}ve 4, Switzerland}

\author{Pablo Jarillo-Herrero}
\email[]{pjarillo@mit.edu}
\affiliation{\MIT}
\maketitle
\newpage

\section{ABA trilayer graphene Hamiltonian}

The Hamiltonian of ABA trilayer graphene can be written as follows~\cite{Dresselhaus2002,Koshino2009},

\begin{equation}
H=\left(
  \begin{array}{cc}
    H_m & \Gamma_D \\
    {\Gamma^T_D} & H_b \\
  \end{array}
\right),
\label{eq_S1}
\end{equation}

The two blocks along the diagonal - $H_m$ and $H_b$ - are the Hamiltonians for the monolayer graphene-like and bilayer graphene-like sectors, which are given by:

\begin{equation}
H_m=\left(
      \begin{array}{cc}
        \Delta_2-\gamma_2/2 & \upsilon\pi^\dagger \\
        \upsilon\pi & \Delta_2-\gamma_5/2+\delta \\
      \end{array}
    \right),
\,\,\,\,\,\,
H_b=\left(
      \begin{array}{cccc}
      \Delta_2+\gamma_2/2             & \sqrt{2}\upsilon_3\pi          & -\sqrt{2}\upsilon_4\pi^\dagger & \upsilon\pi^\dagger \\
      \sqrt{2}\upsilon_3\pi_\dagger     & -2\Delta_2                     & \upsilon\pi                    & -\sqrt{2}\upsilon_4\pi \\
      -\sqrt{2}\upsilon_4\pi          & \upsilon\pi^\dagger            & -2\Delta_2+\delta              & \sqrt{2}\gamma_1 \\
      \upsilon\pi                     & -\sqrt{2}\upsilon_4\pi^\dagger & \sqrt{2}\gamma_1              & \Delta2+\gamma_5/2+\delta \\
      \end{array}
    \right).
\label{eq_S2}
\end{equation}
The off-diagonal block ($\Gamma_D$) is responsible for coupling $H_m$ and $H_b$. The matrix elements of $\Gamma_D$ depend only on the potential difference between the two graphene outer layers $\Done$, as we show with more detail below.

\begin{equation}
    \Gamma_D=\left(
          \begin{array}{cccc}
            \Done & 0 & 0 & 0     \\
                0 & 0 & 0 & \Done \\
          \end{array}
        \right).
\end{equation}

In summary, the ABA Hamiltonian is written as a function of 9 parameters. Six out of the 9 parameters are Slonczewski-Weiss-McClure (SWMcC) parameters ($\gamma_0$,$\gamma_1$,$\gamma_2$,$\gamma_3$,$\gamma_4$,$\gamma_5$), which are related to couplings in the tight binding model\cite{Koshino2009}. Another parameter is $\delta$, which describes the on-site energy difference between A and B sublattice within the same layer. And, finally, the potential differences between layers are characterized by $\Delta_1 = (-e)(U_1 - U_3)/2$ and $\Delta_2 = (-e)(U_1 - 2 U_2 + U_3)/6$.

With application of electric field, $\Gamma_D$ assumes non-trivial values causing hybridization of the TLG subbands\cite{Koshino2009,Campos2012}. Therefore, also the energies of the Landau levels assume values dependent on electric field. Then, to simplify the quantum Hall analysis, we measure magnetoresistance at $\Done = 0$, as we show in Figure 1(c) in the main text and Fig. S1(a).

\section{Landau level spectrum simulations at zero displacement field and fitting procedures}

At zero displacement field, ABA trilayer graphene Hamiltonian gets simplified with $\Gamma_D$ = 0. More precisely, the energies of the Landau levels stay fixed, independent of the charge density\cite{Serbyn2013}. Here, in order to determine SWMcC parameters, we mapped all Landau level crossings shown in Figure 1(c) (main text), Figure S1(a), and then we performed numerical calculations of the energy of the Landau levels as a function of magnetic field (see more details in ref.~\cite{Taychatanapat2011}). Ideally, we should vary the band parameters until the crossing points from our simulations match with the experimental data. However, even with the Hamiltonian simplified by trivial $\Gamma_D$, zero displacement field does not guarantee that $\Dtwo$ also is zero. Indeed, numerical simulations show that the Landau level crossings observed in Fig.~1(c) can not be described by simple numerical solutions with any constant $\Dtwo$. This forces us to invest in a better numerical approach capable to incorporate non-zero values of $\Dtwo$.

Non-zero parameter $\Dtwo$ is allowed by symmetry at $\Done$~=~0, and essentially it occurs if the charge density on the middle layer is different of zero. This is not favorable by electrostatic energy, but there can be a compensating energy gain from the band structure. For instance: the eigenfunction of the TLG Hamiltonian is written in a basis with components $\psi_{A1}$, $\psi_{B1}$, $\psi_{A2}$, $\psi_{B2}$, $\psi_{A3}$, $\psi_{B3}$. If one solution of a specific Landau level has zero values at $\psi_{A1}$, $\psi_{B1}$ and $\psi_{A3}$, $\psi_{B3}$, this indicates that all charge will be added in the middle layer when this hypothetical Landau level gets filled. Therefore, by evaluating the eigenfunctions of the solutions of the Hamiltonian as a function of magnetic field, we estimated how charges get distributed between the three layers at $\Done$~=~0. $\Dtwo$ is calculated using the following expression~\cite{Koshino2009}, $\Dtwo = -2 \pi e^2 d n_2/3 \epsilon_r$, where $n_2$ is the charge density on the central layer, $d$ is the interlayer separation, $e$ is the electron charge and $\epsilon_r$ is the dielectric constant of graphene.  Here, we assume $\epsilon_r$ = 1 and the value of $\Dtwo$ are then calculated for any given filling factor and SWMcC parameters.

\begin{center}
    \begin{tabular}{|c|c|c|c|c|c|c|}
    \multicolumn{7}{c}{Table S1. Tight binding parameters and $\delta$ ranges considered} \\
      \hline
      $\gamma_0$ (eV) & $\gamma_1$ (meV) & $\gamma_2$ (meV) & $\gamma_3$ (meV) & $\gamma_4$ (meV)& $\gamma_5$ (meV)& $\delta$ (meV) \\ \hline
     3.1 & 390 & -25 to 6  & 315 & 40 to 200 & 0 to 70& 0 to 85 \\ \hline

    \end{tabular}
\end{center}

In our simulations, we explore a wide range of values for each of the SWMcC parameter to include all reported possibilities \cite{Dresselhaus2002,Zhang2008b,Taychatanapat2011}. To reduce a number of fitting parameters, we fix the most studied ones like $\gamma_0$ and $\gamma_1$ as $3.1$~eV and $0.39$~eV respectively. The parameter $\gamma_0$ describes coupling between in plane nearst-neighbor ($A_1\leftrightarrow B_1:\gamma_0$)~\cite{Koshino2009,Serbyn2013}, while $\gamma_1$ describes the strongest nearst-layer coupling ($B_1\leftrightarrow A_2$ and $A_2\leftrightarrow B_3:\gamma_1$). The next parameter, $\gamma_3$, describes weaker coupling between A and B sites at consecutive layer ($A_1\leftrightarrow B_2$ and $B_2\leftrightarrow A_3:\gamma3$), and it is responsible for trigonal warping. The value of $\gamma_3$ is also fixed at reported value $0.315$~eV since it affects only higher energy Landau levels, which lie above the energies investigated in this work. We, therefore, change values of the other parameters $\gamma_4$, $\gamma_5$, $\delta$ and $\Dtwo$. The fact that $\Dtwo$ depends on the density of charge, and assumes different values for each filling factor adds a serious computational problem. A variable parameter results in an unaffordable computational time to fit the data. Therefore, we take a simplified approach, described below.

First, we only simulate Landau level crossings at high filling factors ($|\nu| \geq 10$). Since $\Dtwo$ is much smaller than the energy gaps between Landau levels at higher filling factors, we use $\Dtwo = -5$~meV ($\Dtwo^{(\nu=10)}$) for the electron regime and $5$~meV ($\Dtwo^{(\nu=-10)}$) for the hole regime to fit our LL crossing patterns at $|\nu| > 10$. Naturally, a consideration of constant values for the parameter $\Dtwo$ is an approximation and adds imprecision on the values of the SWMcC parameters. Then, secondly, we use these values to check if the simulations can describe lower energy features showed in Fig. S1 (a) with those parameters. In this second verification, we calculate $\Dtwo$ for each filling factor and we incorporate into the Landau level calculations. The acceptable values showed in Table S1 are the ones able to describe both crossings at high energy and lower energy features.

The main feature at low energy (see figure Fig.~S1(a)) is an asymmetrical gap at $|\nu| = 2$. There is a Landau level gap at $\nu = 2$ but not at $\nu = -2$ for $B > 4$~T. We first calculate $\Dtwo$ at $\nu = 2$ and at $-2$ for a given set of SWMcC parameter. Then, we use these values of $\Dtwo$ and the SWMcC parameters to simulate the Landau level energies.  The energy gaps of $\nu = \pm 2$ from the simulation are compared with disorder broadening potential which we estimate to be $\sim$2~meV from the width of LL crossings. We then keep SWMcC parameters which result in the energy gap at $\nu = 2$ much larger than $2$~meV and the energy gap at $\nu = -2$ similar to $2$~meV. Figures S1(b-h) show LL spectrum for the following SWMcC parameters $\gamma_0$= 3.1 eV, $\gamma_1 $= 390 meV, $\gamma_2$= -18 meV, $\gamma_3$= 315 meV, $\gamma_4$= 100 meV, $\gamma_5$= 10 meV and $\delta$= 15 meV with different $\Dtwo$ calculated for different filling factors. Figure S1(f) shows LL spectrum for $\Dtwo = 2.4$~meV calculated for $\nu = -2$. At this value of $\Dtwo$, the energy gap at $\nu = -2$ is similar to the disorder broadening potential which agrees with the fact that we cannot observe the gap at $\nu = -2$ showed in our data.

\begin{figure}
  \includegraphics[width=153mm]{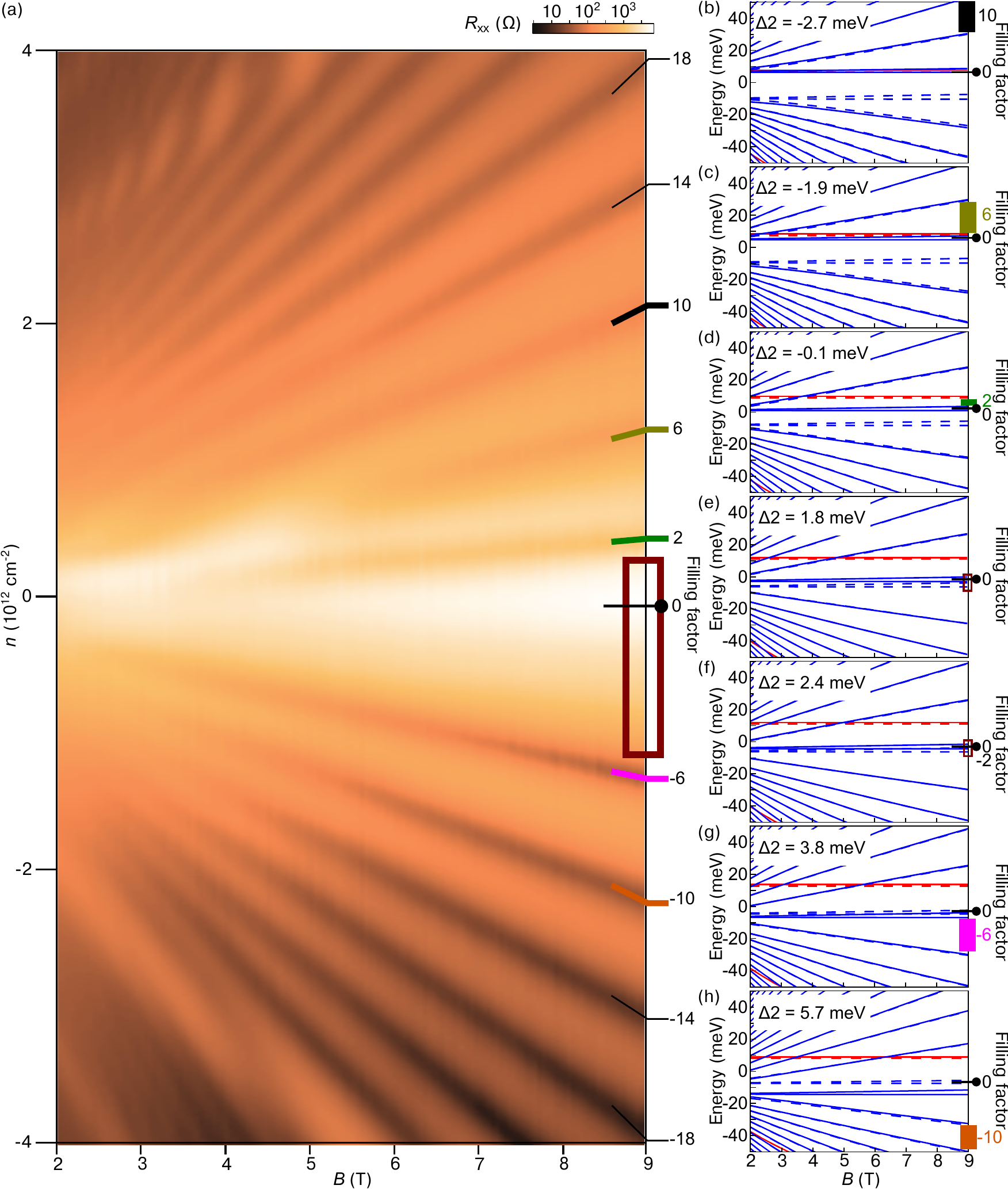}
  \caption{\textbf{Quantum Hall experiment and simulations as a function of $\Dtwo$ at $B$ = 9 T}
\textbf{(a)}, Longitudinal resistance in the quantum Hall regime on sample 1.
\textbf{(b)}, Landau Level simulation at $\nu$ = 10 ($\Dtwo$ = -2.7 meV); LL separation of $\sim$ 23 meV.
\textbf{(c)}, Landau Level simulation at $\nu$ = 6 ($\Dtwo$ = -1.9 meV); LL separation of $\sim$ 22 meV.
\textbf{(d)}, Landau Level simulation at $\nu$ = 2 ($\Dtwo$ = -0.1 meV); LL separation of $\sim$ 5 meV.
\textbf{(e)}, Landau Level simulation at $\nu$ = 0 ($\Dtwo$ = 1.9 meV); LL separation smaller than 1 meV .
\textbf{(f)}, Landau Level simulation at $\nu$ = -2 ($\Dtwo$ = 2.4 meV); LL separation smaller than 2 meV.
\textbf{(g)}, Landau Level simulation at $\nu$ = -6 ($\Dtwo$ = 3.8 meV); LL separation of $\sim$ 23 meV.
\textbf{(h)}, Landau Level simulation at $\nu$ = -10 ($\Dtwo$ = 5.7 meV); LL separation of $\sim$ 23 meV.
  }\label{f1}
\end{figure}

\bibliographystyle{apsrev4-1}